\documentclass[aps,pra,twocolumn,groupedaddress]{revtex4-1}

\usepackage{graphicx}
\usepackage{dcolumn}
\usepackage{bm}
\usepackage{amsmath}
\usepackage{amssymb}
\usepackage{latexsym}
\usepackage{epsfig}
\usepackage{amsbsy}
\usepackage{array}
\usepackage{amssymb}
\usepackage{setspace}
\usepackage{bm}
\usepackage{epstopdf}
\usepackage{subfigure}

\def\sint{\ifmmode{- \!\!\!\!\!\! \int}
    \else{\hbox{$- \!\!\!\! \int \ $}}\fi}

\begin{document}

\title{Asymmetrical interaction induced real spectra and exceptional points in a non-Hermitian Hamiltonian}

\author{Wenlin Li}
\author{Chong Li}
\author{Heshan Song}
\email{hssong@dlut.edu.cn}
\affiliation{School of Physics, Dalian University of Technology, 116024, China}

\date{\today}
\begin{abstract}
Non-Hermitian systems with parity-time symmetry have been developed rapidly and hold great promise for future applications. Unlike most existing works considering the symmetry of the free energy terms (e.g., gain-loss system), in this paper, we report that a realizable non-Hermitian interaction between two quantum resonances can also have a real spectrum after the exceptional point. That phenomenon is similar with that in the gain-loss system so that the non-Hermitian interaction can be an excellent substitute for  quantum gain. Such a non-Hermitian interaction can be realized in designed optomechanics, and we find that its dynamics are in accordance with those of normal gain system as expected. As examples, the phase transition near the exceptional point and the induced chaos in weak nonlinear coupling are shown and analyzed for an intuitive visual. Our results provide a platform for realizing parity-time symmetry devices and studying properties of non-Hermitian quantum mechanics.
\end{abstract} 
\maketitle
\section{Introduction}  
Since Bender \textit{et  al}. noted  that a non-Hermitian Hamiltonian can also have entirely real eigenenergies if the system possesses a combined parity and time-reversal symmetry \cite{1,2}, the studies of non-Hermitian systems with such a parity-time symmetry ($H\neq H^{\dagger}$, $[H,PT]=0$) are developing rapidly in recent years \cite{3,4,5,6,7,8}. As an extension of Hermitian quantum mechanics, besides its new properties in mathematics \cite{9,10,11,12,add8}, the realizing and observing of parity-time ($\mathcal{PT}$)-symmetric non-Hermitian Hamiltonian with existing Hermitian system and technology is also a popular topic in quantum optics. Until now, two representative schemes have been discussed widely as mature simulations of a system with $\mathcal{PT}$-symmetric non-Hermitian Hamiltonian. One of them regards the waveguide as a quantum system \cite{13,14} and similar ideas are also extended into atomic systems in recent years \cite{15,16}. The transmission equation in waveguides is a Schr\"odinger-type equation in which the refractive index corresponds to potential energy in a normal Schr\"odinger equation. With appropriate refractive index, the wave function of a $\mathcal{PT}$-symmetric quantum system can be observed by measuring the optical distribution in the waveguide. Another scheme consider two coupled open quantum resonances with balanced gain and loss \cite{3,4,8,17,18,19,20,21,22}. Compared to the first scheme, gain-loss system does not need other physical quantities to simulate system wave function so that it is also applied widely to enhance quantum information processing, such as optics transmissions \cite{23,24,25}, topological state preparations \cite{19,20,26,27,28}, producing phonon laser \cite{18}, and so on.

With the rapidly developing and future applications of gain-loss $\mathcal{PT}$-symmetric systems, a defect is gradually exposed, that is, gain is not a spontaneous concept in traditional quantum mechanics \cite{29,30,31}. For a single gain system, the non-Hermitian corresponds to a non-conservation probability so that it is strict only in the semi-classical level. The application of $\mathcal{PT}$-symmetric system has been restricted seriously in the quantum regime because most previous works ignored the extra noises caused by gain \cite{17,19,20,25,32,33,34,35}. Unfortunately, existing experimental technique for gain, for example ion doping, is hard to find corresponding complete quantum description.

The aim of this work is to address this problem, in particular, we consider a non-Hermitian interaction \cite{add8,add9,add10} between two quantum resonances. By diagonalization of effective Hamiltonian, we find that the corresponding system energy spectrum is similar with that in gain-loss $\mathcal{PT}$-symmetric system \cite{36}. The energy spectrum can be real even though the system corresponds to a non-Hermitian effective Hamiltonian. Moreover, there exists an exceptional point (EP) and the degenerate real parts (splitting imaginary part)  of the spectrum will become splitting (degenerate) once some key parameters pass the EP \cite{36}.  In gain-loss $\mathcal{PT}$-symmetric system, the EP is a dividing line of $\mathcal{PT}$SP and $\mathcal{PT}$BP. In fact, the most exciting highlight of $\mathcal{PT}$-symmetric system is the presence of EP and around it a series of fascinating phenomena exist, such as $\mathcal{PT}$ phase transitions \cite{36,37,38,39}, induced chaoses \cite{17,40} and so on. The moral is that our non-Hermitian system can be good replacement for  $\mathcal{PT}$-symmetric system and it can avoid introducing any gains into the system.

For discussing and observing dynamic properties of the non-Hermitian interaction, we also design a scheme to realize it in optomechanical systems. With this platform,  the dynamic behaviors caused by non-Hermitian interaction are  compared with those of gain-loss $\mathcal{PT}$-symmetric system. Based on our simulation results, it can be found that some characteristics of gain system, for example phase transition near the EP and gain induced chaos, can also appear with the help of non-Hermitian interaction. 
 
We organize this paper as follows: in Sec. \ref{model}, we analyze the energy spectrum  of the system with non-Hermitian interaction. In Sec. \ref{Non-Hermitian interaction in an  Optomechanical system}, we propose a feasible scheme to realize the ideal model in Sec. \ref{model}. In Sec. \ref{System dynamics with Non-Hermitian interaction}, we present the dynamic properties caused by the non-Hermitian interaction, including the phase transition near the EP of a linear system in Sec. \ref{Phase transition before and after EP} and induced chaos in nonlinear system in Sec. \ref{Non-Hermitian interaction induced Chaos}. Finally, a feasibility analysis and a summary are given in Sec. \ref{discussion and conclusion}.
\section{energy spectrum of system with non-Hermitian interaction}  
\label{model}
\begin{figure}[]
\centering
\includegraphics[width=3.3in]{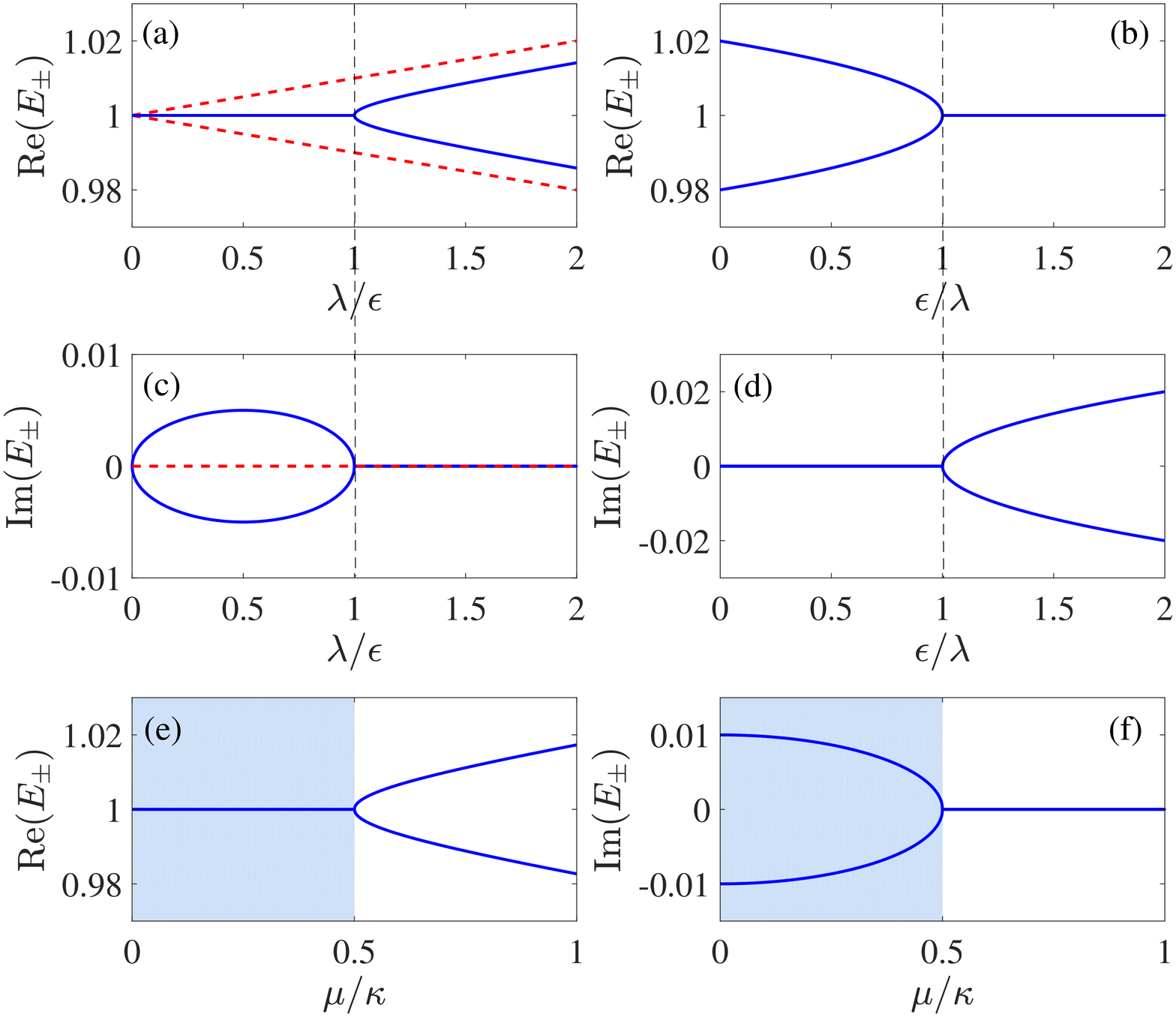}  
\caption{Energy spectra of non-Hermitian interaction system ((a)-(d)) and gain-loss $\mathcal{PT}$-symmetric system ((e) and (f)). The red dotted lines in (a) and (c) are the normal mode splitting corresponding to Hermitian interaction. The blue areas in (e) and (f) correspond to $\mathcal{PT}$BP and white areas are $\mathcal{PT}$SP. Here $\omega=1$ are set as an unit and the other parameters are $\epsilon=0.01$ in (a, c),  $\lambda=0.02$ in (b, d) and $\kappa=0.01$ in (e, f). 
\label{fig:1}}
\end{figure}

We firstly consider a normal Hermitian Hamiltonian with a beam splitter (BS) energy transformation between two quantum oscillators:
\begin{equation}
\begin{split}
H=\omega_cc^\dagger c+\omega_dd^\dagger d+\mu c^\dagger d+\mu^{*} d^\dagger c.
\end{split}
\label{eq:gelHamilton}
\end{equation}
The Heisenberg equations corresponding to this system are  
\begin{equation}
\begin{split}
&\dot{c}=-i\omega_c c-i\mu d\\
&\dot{d}=-i\omega_d d-i\mu^{*} c,
\end{split}
\label{eq:Heisenberg}
\end{equation}
and they can be rewritten in a compact matrix form as $i\dot{\vert\psi\rangle}=H_{\text{eff}}\vert\psi\rangle$ \cite{37}, which is a Schr\"odinger-like equation with $\vert\psi\rangle=(c,d)^{T}$ and the effective Hamiltonian
\begin{equation}
\begin{split}
H_{\text{eff}}=
\begin{pmatrix}
\omega_c &\mu\\
\mu^{*}&\omega_d
\end{pmatrix}.
\end{split}
\label{eq:effective Hamiltonian}
\end{equation}
It is apparent that $H_{\text{eff}}$ is Hermitian and owns real spectrum. Once $\omega_c=\omega_d=\omega$, its energy spectra exhibit a normal mode splitting as shown in Fig. \ref{fig:1}(a) and (c), and there are no effective gain and dissipation in this system.

The Hermiticity of the quantum mechanics ensures that the nondiagonal elements of the effective Hamiltonian should be conjugated. However, if we consider the coupling intensities between the two systems are asymmetrical, for example there is a constant deviation between two nondiagonal elements, then the effective Hamiltonian becomes:
\begin{equation}
\begin{split}
H_{\text{eff}}=
\begin{pmatrix}
\omega &i\lambda\\
-i(\lambda-\epsilon)&\omega
\end{pmatrix}.
\end{split}
\label{eq:iamgHamilton}
\end{equation}
where $\lambda,\epsilon\in\mathcal{R}$. The nondiagonal elements $i\lambda$ and $-i\lambda$ in Eq. \eqref{eq:iamgHamilton} can be obtained by setting $\mu=i\lambda$ in Eq. \eqref{eq:effective Hamiltonian} and the whole Hamiltonian will be Hermitian if and only if $\epsilon=0$. The energy spectra of Eq. \eqref{eq:iamgHamilton} are 
\begin{equation}
\begin{split}
E_{\pm}=\omega\pm\sqrt{\lambda^2-\epsilon\lambda},
\end{split}
\label{eq:eigvalue1}
\end{equation}
and one can see intuitively that there exists an exceptional point (EP) at $\lambda=\epsilon$. Once $\lambda\geq \epsilon$, the second term will have no contribution to the imaginary parts of the eigenvalues and $E_{\pm}$ are still real even though $H_{\text{eff}}$ is no longer Hermitian. Correspondingly, $\lambda<\epsilon$ makes the real parts of eigenvalues degenerate  but the imaginary parts be different. Note that $E_{+}$ can be greater than $0$ even though there is no additional gain in the system. In Fig. \ref{fig:1}(a)-(d), we plot $E_{\pm}$ with varied $\lambda$ ($\epsilon$) but fixed $\omega$ and $\epsilon$ ($\lambda$) to show the different spectra before and after EP.

We note that this phenomenon is similar to that presented by  a $\mathcal{PT}$-symmetric system with  non-Hermitian diagonal terms corresponding to gain and dissipation, respectively. The non-Hermitian Hamiltonian of such a system is \cite{17}
\begin{equation}
\begin{split}
H=\left(\omega+i\dfrac{\kappa}{2}\right)c^\dagger c+\left(\omega-i\dfrac{\kappa}{2}\right)d^\dagger d+\mu(c^\dagger d+ d^\dagger c),
\end{split}
\label{eq:gainHamilton}
\end{equation}
with energy spectra $E'_{\pm}=\omega\pm\sqrt{\mu^2-\kappa^2/4}$ and the corresponding EP is $\mu=\kappa/2$. It determines whether the coupling between two resonances is strong enough to transfer the gained energy to the dissipate resonance and to eliminate the effective dissipation \cite{17,36,37}. As contrast, we also plot $E'_{\pm}$ in Fig. \ref{fig:1}(e) and (f) to show the similar degeneracy and bifurcation of the two cases. Such a similar spectrum allows us to expect the system can have similar dynamics with  $\mathcal{PT}$-symmetric system, and we will discuss its corresponding dynamics in the following sections.
\begin{figure}[]
\centering
\includegraphics[width=3.3in]{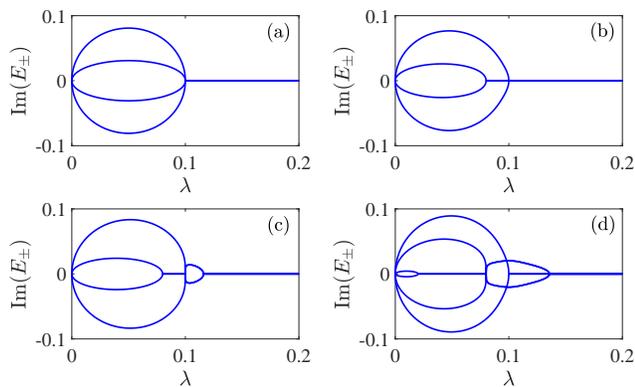}  
\caption{Imaginary parts of energy spectra corresponding to $N=4$ (a)-(c) and $N=6$ (d). In (a), (b) and (c), $(\omega,\epsilon_1,\epsilon_2,\epsilon_3)$ are set as $(1,0.05,0.05,0.05)$, $(1,0.05,0.05,0.04)$ and $(1,0.05,0.06,0.04)$, respectively. In (d),  $(\omega,\epsilon_1,...,\epsilon_5)=(1,0.05,0.06,0.04,0.07,0.01)$. 
\label{fig:add}}
\end{figure}

At the end of this part, the existence of multiple exceptional points \cite{add3} is clarified finally with the increasing cavity number $N$. The effective Hamiltonian for a one-dimensional coupled oscillator-chain with asymmetrical interactions is 
\begin{equation}
\begin{split}
H_{\text{eff}}=
\begin{pmatrix}
\omega &i\lambda &0&0&...&0\\
-i\upsilon_1&\omega &i\lambda&0 &...&0\\
0&-i\upsilon_2&\omega &i\lambda&... &0\\
...&...&....&... &...&...\\
0&0&0&... &-i\upsilon_{N-1}&\omega
\end{pmatrix},
\end{split}
\label{eq:effective Hamiltonian multi}
\end{equation} 
with $\upsilon_i=\lambda-\epsilon_i$. It is perceivable that the spectra and supermodes of the matrix have regular patterns with the increase of the cavity number $N$. We plot their performances in Fig. \ref{fig:add} and they always appear in pairs of opposite signs. Once $\epsilon_i$ are unequal to each other, corresponding exceptional points loss their degeneracy and  multiple exceptional point phenomenon appears. Therefore the dynamics across the different exceptional points can also observed in our model. 

\section{Non-Hermitian interaction in an  Optomechanical system}
\label{Non-Hermitian interaction in an  Optomechanical system}
Now we firstly consider how to realize non-Hermitian interaction in an actual physics system. The key issue of this aim is to get a unidirectional coupling between two systems, which is forbidden in general closed quantum systems. However, for open quantum systems, the leakage information from one system can be thought as an input of another system \cite{41,42}. Under the Markovian approximation, this process can be regarded as a non-Hermitian interaction and this scheme has been applied in QIP \cite{43,44,45,46,47,48}. 

The above non-Hermitian interaction is induced by the cavity leakage, so that the corresponding effective Hamiltonian will contain non-Hermitian dissipation terms, inevitably. In order to eliminate it, spontaneous gain in the system is also necessary for a perfect realization of Hamiltonian Eq. \eqref{eq:iamgHamilton}. In optomechanics, the spontaneous gain is relatively simple to implement. Because the radiation pressure interaction $H_I=-g_0a^{\dagger}a(b^{\dagger}+b)$ is unchanged under the frame rotation $U=\exp(-i\omega_Da^{\dagger}at)$ \cite{49, 50}, one can choose to heat or cool system by adjusting appropriate detuning before a linearization \cite{51,add1}. Therefore, optomechanics are selected as the platform of our scheme. 
\begin{figure}[]
\centering
\includegraphics[width=3.3in]{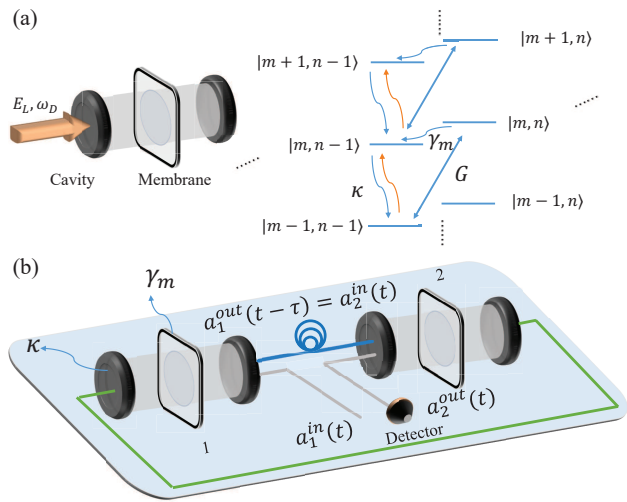}  
\caption{(a): A schematic illustration of an optomechanics. On the right is the level diagram of the linearized Hamiltonian, and $\vert m,n\rangle$ denotes the state of $m$ photons and $n$ phonons in the displaced frame. (b): The realization of a non-Hermitian interaction in optomechanics. The green line denotes a BS interaction with tunneling strength $\lambda$ and the blue line refers to an unidirectional fiber.
\label{fig:2}}
\end{figure}

As shown in Fig. \ref{fig:2}(a), such a normal optomechanical system can be linearized under strong driving condition and the weak coupling regime ($E/\omega_m\gg 1$, $g/\kappa\ll 1$) \cite{50}. We note that there is an optical heating mechanism with blue detuning and larger oscillator dissipation ($\omega_d>\omega_c$, $\gamma_m\gg\kappa$), that is, the linearized  coupling between optical mode and mechanical mode stimulates the system from $\vert m,n\rangle$ to $\vert m+1,n+1\rangle$, and the latter will lose a phonon and decay to $\vert m+1,n\rangle$. These two processes are equivalent to heat the optical mode ($\vert m\rangle\rightarrow \vert m+1\rangle$) after eliminating the mechanical mode (orange arrow in Fig. \ref{fig:2}(a)), and it can balance normal cavity dissipation effect that decays the cavity field from $\vert m+1\rangle$ to $\vert m\rangle$.

Now we discuss the details in mathematics. After a rotating wave approximation (RWA), the linearized Hamiltonian 
\begin{equation}
\begin{split}
H'_l=-\Delta a^\dagger a+\omega_m  b^\dagger b+(G^*a^\dagger+Ga)(b^\dagger +b).
\end{split}
\label{eq:omelinerHamiltonyuanshi}
\end{equation}
corresponding to an optomechanical system with blue detuning can be simplified as \cite{49}:
\begin{equation}
\begin{split}
H_l=-\Delta a^\dagger a+\omega_m  b^\dagger b+G^*a^\dagger b^\dagger+Gab.
\end{split}
\label{eq:omelinerHamilton}
\end{equation}
and the total Hamiltonian of the system shown in Fig. \ref{fig:2}(b) is 
\begin{equation}
\begin{split}
H=&\sum_{j=1,2}\left(-\Delta_j a^\dagger_j a_j+\omega_m  b^\dagger_j b_j+G^*a_j^\dagger b_j^\dagger+Ga_jb_j\right)\\
&+i\lambda a^\dagger_1a_2-i\lambda a_1a^\dagger_2.
\end{split}
\label{eq:totalHamilton}
\end{equation}
The dynamics of the system including dissipation and decoherence effects can then be described with the following quantum Langevin equations \cite{52}:
\begin{equation}
\begin{split}
&\dot{a}_j=\left(i\Delta_j-\dfrac{\kappa}{2}\right)a_j-iG^{*}b^\dagger_j-(-1)^{j}\lambda a_{3-j}+\sqrt{\kappa}a^{in}_j\\
&\dot{b}_j=\left(-i\omega_m-\dfrac{\gamma_m}{2}\right)b_j-iG^{*}a^\dagger_j+\sqrt{\gamma_m}b^{in}_j.
\end{split}
\label{eq:qle}
\end{equation}
In addition to the interaction term $i\lambda a^\dagger_1a_2-i\lambda a_1a^\dagger_2$, here we also consider an unidirectional coupling between the cavities, that is, the output field of the first cavity constitutes the input for the second cavity with an appropriate time delay. However, the first cavity is decoupled from the second one, which implies that the input and output operators should satisfy: $a^{out}_{j}(t)=\sqrt{\kappa}a_j(t)-a^{in}_{j}(t)$ and $a^{in}_{2}(t)=a^{out}_{1}(t-\tau)$ with constant $\tau$ relating to retardation in the propagation between the mirrors \cite{42,48}. The input-output relation can be rewritten as
\begin{equation}
\begin{split}
a^{in}_{2}(t)=\sqrt{\kappa}a_1(t-\tau)-a^{in}_{1}(t-\tau).
\end{split}
\label{eq:guangxianshuru}
\end{equation}
Note that $a^{in}_1$ is the vacuum input with $\langle a^{in}_1\rangle=0$ \cite{add6}, so we have a set of differential equations by taking the mean values of all the operators into Eq. \eqref{eq:qle}:
\begin{equation}
\begin{split}
&\dot{\alpha}_1=\left(i\Delta_1-\dfrac{\kappa}{2}\right)\alpha_1-iG^{*}\beta^*_1+\lambda \alpha_2\\
&\dot{\beta}_1=\left(-i\omega_m-\dfrac{\gamma_m}{2}\right)\beta_1-iG^{*}\alpha^*_1\\
&\dot{\alpha}_2=\left(i\Delta_2-\dfrac{\kappa}{2}\right)\alpha_2-iG^{*}\beta^*_2-\lambda \alpha_1+\kappa\alpha_1\\
&\dot{\beta}_2=\left(-i\omega_m-\dfrac{\gamma_m}{2}\right)\beta_2-iG^{*}\alpha^*_2.
\end{split}
\label{eq:meavalueqle}
\end{equation}
In this expression, $\alpha_j=\langle a_j\rangle$, $\beta_j=\langle b_j\rangle$ and the short fiber approximation is applied so that $\tau\rightarrow 0$ \cite{42}. 

After eliminating the oscillator modes under the parameter conditions $\omega_m-\Delta_j\gg \vert G\vert$, $\alpha_j(0)\gg\beta_j(0)$ and $\gamma_m\gg\kappa$ \cite{add4,add5}, the remainder of the differential equations are: 
\begin{equation}
\begin{split}
&\dot{\alpha}_1=\left(i\Delta_{\text{eff}1}-\dfrac{\Gamma}{2}\right)\alpha_1+\lambda \alpha_2\\
&\dot{\alpha}_2=\left(i\Delta_{\text{eff}2}-\dfrac{\Gamma}{2}\right)\alpha_2-\lambda \alpha_1+\kappa\alpha_1,
\end{split}
\label{eq:effectmeavalueqle}
\end{equation}
with the effective frequency
\begin{equation}
\begin{split}
\Delta_{\text{eff}j}=\Delta_j-\dfrac{4(\Delta_j-\omega_m)\vert G\vert^2}{4(\Delta_j-\omega_m)^2+\gamma_m^2},
\end{split}
\label{eq:effectxishu}
\end{equation}
and dissipation
\begin{equation}
\begin{split}
\Gamma=\kappa-\dfrac{4\gamma_m \vert G\vert^2}{4(\Delta_j-\omega_m)^2+\gamma_m^2}.
\end{split}
\label{eq:effectxishu}
\end{equation}
The effective dissipation can be balanced completely ($\Gamma=0$) with $\vert G \vert^2=\kappa[4(\Delta_j-\omega_m)^2+\gamma_m^2]/4\gamma_m$. Substituting it to effective frequency, we have:
\begin{equation}
\begin{split}
&\Delta_{\text{eff}j}=\Delta_j-\dfrac{\kappa(\Delta_j-\omega_m)}{\gamma_m}, 
\end{split}
\label{eq:effectfrequencie}
\end{equation}
and finally we can obtain 
\begin{equation}
\begin{split}
&\dot{\alpha}_1=i\Delta_{\text{eff}1}\alpha_1+\lambda \alpha_2\\
&\dot{\alpha}_2=i\Delta_{\text{eff}2}\alpha_2-(\lambda-\kappa)\alpha_1,
\end{split}
\label{eq:effectmeavalueqlefinal}
\end{equation}
to simulate Eq. \eqref{eq:iamgHamilton}. Note that there is a sign difference between the eigenfrequencies of Eq. \eqref{eq:iamgHamilton} and Eq. \eqref{eq:effectmeavalueqlefinal}, on the other words, Eq. \eqref{eq:effectmeavalueqlefinal} can be obtained by setting $\omega$ with negative value. But this difference does not affect the phase transition and EP caused by non-Hermitian interactions. 
\begin{figure}[]
\centering
\includegraphics[width=3.3in]{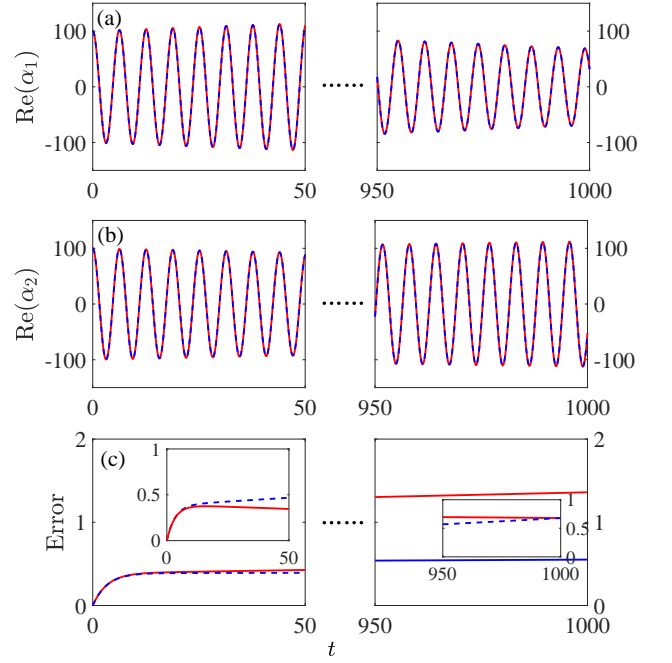}  
\caption{(a) and (b):  Evolutions of $\alpha_1$  and $\alpha_2$ based on accuracy Langevin equations (blue dotted lines) and effective Langevin equations (red solid lines).  (c): Errors between two results, here, the blue dotted line denotes the error of $\alpha_1$ and red solid line is the error of $\alpha_2$. In simulations of (a) and (b), we set $\Delta_j=1$, $\omega_m=1.05$, $G=0.015$, $\kappa=0.0015$, $\lambda=2\kappa=0.003$, $\gamma_m=0.584$ so that $\Gamma\simeq 2.8\times 10^{-6}\simeq 0$, $\alpha_j(0)=100$ and $\alpha_j(0)/\beta_j(0)=20\gg 1$. In (c), the main figure is $\lambda=0.8\kappa=0.0012$ and the inset is $\lambda=2\kappa=0.003$.
\label{fig:3}}
\end{figure}

In Fig. \ref{fig:3}(a) and (b), we show the contrasts between the accuracy Langevin equations \eqref{eq:meavalueqle} and effective Langevin equations \eqref{eq:effectmeavalueqle} and it can be known that both dynamic equations exhibit the consistent evolutions under our approximate condition. For a more quantitative expression, in Fig. \ref{fig:3}(c), we calculate the error $\vert\alpha_j-\alpha_j'\vert$ to show the approximation validity \cite{add2}. Here, the parameters are set as $\lambda=0.8\kappa$ (main figure) and $\lambda=2\kappa$ (inset figure), corresponding to before and after the EP, respectively. It illustrates  the errors corresponding to the two cases are both smaller than $2$, that is, a $2/\%$ error compared to the initial state even in the long-time regime. Therefore, we think the approximation is tenable and we will utilize Eq. \eqref{eq:effectmeavalueqlefinal} directly in the following discussions.

In addition to the non-Hermitian Hamiltonian, we also want to emphasize here that all the quantum properties of the system, such as fluctuation and non-local correlation, can also be  obtained because we already have the quantum Langevin equations \eqref{eq:qle}. In especial, the covariance matrix can be calculated easily by using of coefficient matrix \cite{53} or stochastic Langevin equation method \cite{54} once  the system state is restricted in the form of Gaussian. Moreover, such a system can also be solved in Schr\"odinger picture with a strictly derived Markovian quantum master equation \cite{55}.
\section{System dynamics with Non-Hermitian interaction}
\label{System dynamics with Non-Hermitian interaction}
To investigate the different dynamical behaviors before and after the EP, we will consider two different models corresponding respectively to two linear coupled resonators with and without nonlineared Hamiltonians to compare the non-Hermitian system and normal gain-loss system with $\mathcal{PT}$-symmetry or  broken $\mathcal{PT}$- symmetry.
\subsection{Phase transition before and after EP}
\label{Phase transition before and after EP}
\begin{figure}[]
\centering
\includegraphics[width=3.3in]{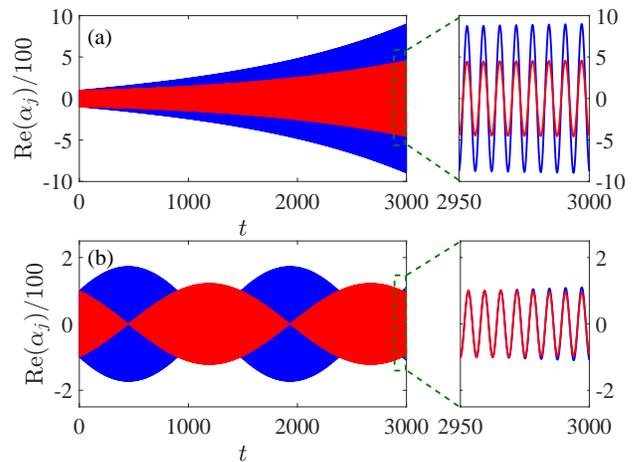}  
\caption{Dynamical behaviors of the two fields. Blue (red) lines denote the evolutions of Re$(\alpha_1)$ (Re$(\alpha_2)$). (a) and (b) correspond to $\lambda$ before and after the EP, that is, $\lambda=0.8\kappa=0.0012$ and $\lambda=2\kappa=0.003$ with $\kappa=0.0015$. Here we set $\Delta_{\text{eff}1,2}=1$, $\Gamma=0$ and the other parameters are the same with those in Fig. \ref{fig:3}.
\label{fig:4}}
\end{figure}

As given in Ref. \cite{37}, the energy transition between $\mathcal{PT}$SP and $\mathcal{PT}$BP is always a key feature to distinguish the two dynamical phases. Therefore, we first consider two pure linear oscillation systems or optical fields whose dynamics should obey Eq. \eqref{eq:effectmeavalueqle} once there exists non-Hermitian interaction. First of all, we simulate the dynamical behavior of the two fields by solving Eq. \eqref{eq:effectmeavalueqle} numerically with the parameters $\Delta_{\text{eff}1,2}=1$ and $\Gamma=0$ and the results are shown in Fig. \ref{fig:4}. What needs to be explained is that the parameters mentioned above can be always satisfied by adjusting original parameters with the relation in Eq. \eqref{eq:effectxishu}.

\begin{figure}[]
\centering
\includegraphics[width=3.4in]{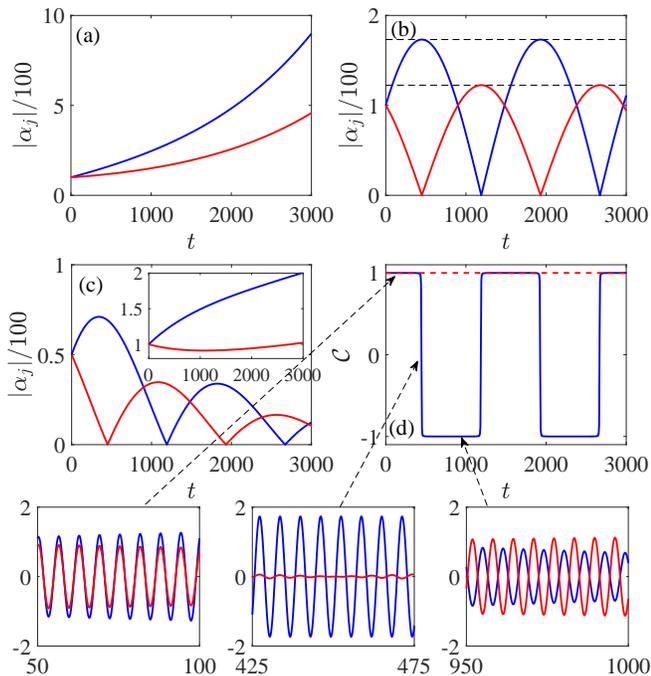}  
\caption{(a,b,c): Evolutions of $\vert \alpha_1\vert$ and $\vert \alpha_2\vert$. (d): Pearson's factors of Re$(\alpha_1)$ and Re $(\alpha_2)$. Here, (a), the inset in (c) and the red dash line in (d) correspond to the case $\lambda$ before the EP, respectively. (b), main figure in (c) and blue solid line in (d) are the case $\lambda$ after the EP. The parameters of (a)-(d) are the same with those in Fig. \ref{fig:4} except for $\Gamma=0.001$ in (c). In (d), the time window is set as $\Delta t=10$. 
\label{fig:5}}
\end{figure}

Through comparison we find that both the amplitudes of two fields present significant increase when $\lambda=0.8\kappa<\kappa$, that is, before the EP. Contrarily, the energies of the two fields will take on periodic exchange between each other once $\lambda$ passes the EP ($\lambda=2\kappa$), and both two amplitudes are confined to limited boundaries rather than continuous exponential amplifications like Fig. \ref{fig:4}(a). In Fig. \ref{fig:5}, we plot amplitudes of two cavity fields for a more intuitive description. Besides, as shown in insets on the bottom  of Fig. \ref{fig:5}, one can observe that the two fields are phase synchronization \cite{59} when $\lambda<\kappa$, contrarily, there has a synchronization anti-synchronization crossover once $\lambda$ passes the EP. 

In mathematics, the above characteristics are obvious according to the analytic solutions of the cavity fields:
\begin{equation}
\begin{split}
&{\alpha}_1=\dfrac{(e^{-iE_+t}+e^{-iE_-t})}{2}c_1+\dfrac{i\zeta(e^{-iE_+t}-e^{-iE_-t})}{2}c_2\\
&{\alpha}_2=\dfrac{-i(e^{-iE_+t}-e^{-iE_-t})}{2\zeta}c_1+\dfrac{(e^{-iE_+t}+e^{-iE_-t})}{2}c_2,
\end{split}
\label{eq:analytic solutions}
\end{equation}
where $c_1=\alpha_1(0)$, $c_2=\alpha_2(0)$, $\zeta=\sqrt{\lambda}/\sqrt{\lambda-\kappa}$ and $E_{\pm}=-\Delta\pm\sqrt{\lambda(\lambda-\kappa)}$ under the definition $\Delta_{\text{eff}1,2}=\Delta$. These solutions can be solved with the help of the energy spectrum, then the biorthogonal basis \cite{56,57} of the effective Hamiltonian can be gained as:  
\begin{equation}
\begin{split}
&\vert\phi_1\rangle=
\begin{pmatrix}
\frac{i\sqrt{\lambda(\lambda-\kappa)}}{\lambda-\kappa} \\
1
\end{pmatrix},\,\,\,\,\,\,
\vert\phi_2\rangle=
\begin{pmatrix}
\frac{-i\sqrt{\lambda(\lambda-\kappa)}}{\lambda-\kappa} \\
1
\end{pmatrix},
\end{split}
\label{eq: biorthogonal basis}
\end{equation}
and 
\begin{equation}
\begin{split}
&\langle\psi_1\vert=
\begin{pmatrix}
\frac{-i\sqrt{\lambda(\lambda-\kappa)}}{\lambda}, &
1
\end{pmatrix},\,\,\,\,\,\,
\langle\psi_2\vert=
\begin{pmatrix}
\frac{i\sqrt{\lambda(\lambda-\kappa)}}{\lambda}, &
1
\end{pmatrix},
\end{split}
\label{eq: biorthogonal basis}
\end{equation}
under the number-basis $\vert 1\rangle=(1,0)^{T}$ and $\vert 2\rangle=(0,1)^{T}$, and one can verify easily that they satisfy the biorthogonal relations
\begin{equation}
\begin{split}
\langle\psi_i\vert \phi_j\rangle=2\delta_{ij},
\end{split}
\label{eq: zhengjiaoguiyi}
\end{equation}
and generalized completeness relations \cite{37,57}
\begin{equation}
\begin{split}
\sum_i\dfrac{\vert \phi_i\rangle\langle\psi_i\vert}{\langle\psi_i\vert \phi_i\rangle}=I_{2\times 2}.
\end{split}
\label{eq: zhengjiaoguiyi}
\end{equation}
Then the initial state $\vert\Psi (0)\rangle=\sum_n c_n\vert n\rangle$ can be rewritten in the biorthogonal basis, and we have:
\begin{equation}
\begin{split}
\vert\Psi (t)\rangle&=e^{-iH_{\text{eff}}t}\vert\Psi (0)\rangle\\
&=e^{-iH_{\text{eff}}t}\sum_n c_n \sum_i\dfrac{\vert \phi_i\rangle\langle\psi_i\vert}{\langle\psi_i\vert \phi_i\rangle} \vert n\rangle\\
&=\sum_n \sum_i \dfrac{ c_n\langle\psi_i\vert n\rangle}{\langle\psi_i\vert \phi_i\rangle}e^{-iE_it} \vert \phi_i\rangle,
\end{split}
\label{eq: motai}
\end{equation}
where $E_1=E_+$ and $E_2=E_-$. Eq. \eqref{eq:analytic solutions} can be obtained by $\alpha_1(t)=\langle 1\vert \Psi (t)\rangle$ and $\alpha_2(t)=\langle 2\vert \Psi (t)\rangle$.

Physically, when $\lambda$ is before the EP, the supermode having positive imaginary part eigenvalue (see the spectrum in Fig. \ref{fig:1}) is just equal to getting a gain so that the system finally shows increasing amplitude. If $\lambda$ passes the EP, the degenerate zeroed effective decay rates make the energy of the system be limited in definite boundaries. Moreover, the fields become two beat-frequency fields because the two supermodes present a mode splitting in their real parts, and the BS interaction controls them to be synchronous or anti-synchronous. In Fig. \ref{fig:5}(d), the synchronization is measured by Pearson's factor \cite{54,55,58}
\begin{equation}
\begin{split}
\mathcal{C}_{f,g}(t,\Delta t)=\dfrac{\overline{\delta f\delta g}}{\sqrt{\overline{\delta f ^2}\times\overline{\delta g ^2}}},
\end{split}
\label{eq:Pearson's factor}
\end{equation}
from which the synchronization and anti-synchronization correspond to $1$ and $-1$ and this crossover can be shown.

We want to emphasize here that both the increasing amplitudes and beat-frequency energy exchange perform the same dynamic behaviors corresponding to $\mathcal{PT}$BP and $\mathcal{PT}$SP in a gain-loss $\mathcal{PT}$-symmetric system. Moreover, the inset of Fig. \ref{fig:5}(c) shows that amplitude of a cavity field increases with an initial decrease once the cavity dissipations are not balanced perfectly. This is also accord with the characteristics of gain-loss system \cite{37}. Because all the properties are possessed, in this linear model, the non-Hermitian interaction Hamiltonian can do a perfect replacement for $\mathcal{PT}$-symmetric system even though we do not add any extra gain into the system. 

\subsection{Non-Hermitian interaction induced Chaos}
\label{Non-Hermitian interaction induced Chaos}
In recent years, gain-loss $\mathcal{PT}$-symmetric systems have also been applied in nonlinear systems of quantum optics. For examples, Ref. \cite{32} has proved that the gain can enhance nonlinear coupling of a cavity-QED system and L\"u \textit{et al}. noticed that $\mathcal{PT}$BP can even induce chaos, a strong nonlinear effect in optomechanics \cite{17}. As a contrast, here we also discuss the chaos phenomenon induced by the non-Hermitian interaction. 
\begin{figure}[]
\centering
\includegraphics[width=3.0in]{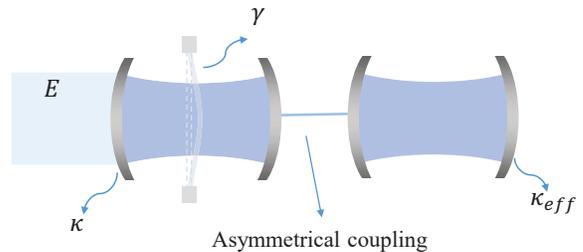}  
\caption{Schematic illustration of optomechanics for realizing non-Hermitian interaction induced chaos. Two optomechanics are coupled by the asymmetrical coupling. The second cavity (right one) has effective frequency and cavity dissipation because its oscillator is eliminated.
\label{fig:6}}
\end{figure}

We consider two optomechanics coupled by the asymmetrical coupling discussed above, as shown in Fig. \ref{fig:6}. The first optomechanics consists of a oscillator whose eigenfrequency and decay are $\omega_{cm}$ and $\gamma$ respectively, and it couples the cavity field with single-photon coupling intensity $g_0$. The oscillator of the second optomechanics (right one) has been eliminated after linearization. The effective semi-classical Langevin equations corresponding to this system are:
\begin{equation}
\begin{split}
&\dot{q}=\omega_{cm} p\\
&\dot{p}=-\omega_{cm} q-{\gamma}p+g_0 \vert\alpha_1\vert^2\\
&\dot{\alpha}_1=\left(i\Delta_c-\dfrac{\kappa_c}{2}\right)\alpha_1+ig_0\alpha_1q+E+\lambda_c\alpha_2\\
&\dot{\alpha}_2=\left(i\Delta_{c\text{eff}}-\dfrac{\kappa_{c\text{eff}}}{2}\right)\alpha_2-\lambda_c\alpha_1+\kappa_c\alpha_1,
\end{split}
\label{eq:chaosqle}
\end{equation}
where $\Delta_{c\text{eff}}$ and $\kappa_{c\text{eff}}$ are adjustable effective frequency and cavity dissipation after eliminating the large detuning oscillator in the second cavity. Since the linearization has been applied on the second cavity, it requires $\alpha_2(0)\gg\alpha_1(0)$. 

\begin{figure}[]
\centering
\includegraphics[width=3.3in]{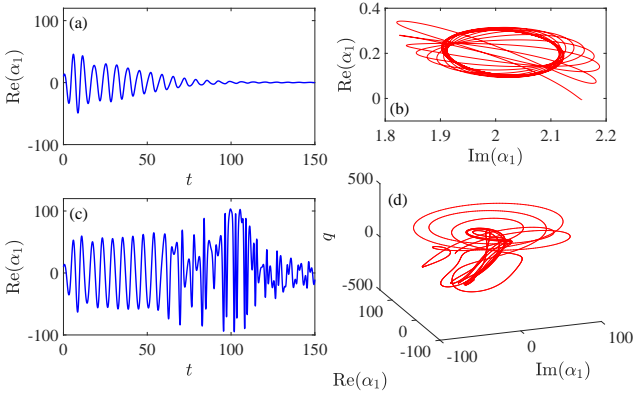}  
\caption{Periodic (a,b) and chaotic (c,d) evolutions of Re$(\alpha_1)$ and their corresponding phase diagrams. In our simulations, (a, b) and (c, d) are the cases of Hermitian and non-Hermitian interaction respectively, that is, the last term in the last equation of Eq. \eqref{eq:chaosqle} is non-existent or existent. Here $\omega_{cm}=1$ is adopted as an unit and other parameters are $g_0=0.01$, $\kappa_c=0.2$, $E=2$, $\Delta_c=\Delta_{c\text{eff}}=1$, $\kappa_{c\text{eff}}=0$, $\gamma=0.038/46\sim 8.2609\times10^{-4}$, $\lambda_c=\kappa_c/2=0.1$, $\alpha_2(0)/\alpha_1(0)=10$ with $\alpha_2(0)=100$ and $q(0)=p(0)=0$. 
\label{fig:7}}
\end{figure}
\begin{figure}[]
\centering
\includegraphics[width=3.3in]{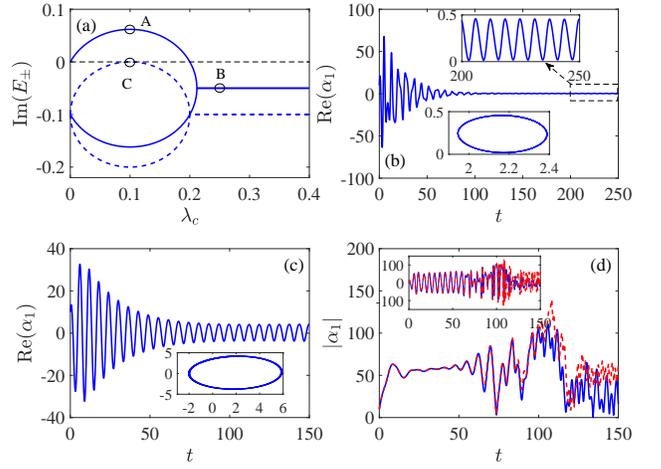}  
\caption{(a): Energy spectra of two cavities after ignoring optomechanical coupling ($g_0=0$). The solid line denotes the case $\kappa_{c\text{eff}}=0$ and the dashed line means the case $\kappa_{c\text{eff}}=\kappa_c$. (b, c): Evolutions of Re$(\alpha_1)$ and their corresponding phase diagrams while $\lambda$ passes the EP ($\lambda_c=0.25$, $\kappa_{c\text{eff}}=0)$ and there is no heating mechanism ($\kappa_{c\text{eff}}=\kappa_c$, $\lambda_c=0.1$). (d): Chaoses with different initial states (blue: $\alpha_1=10$, red: $\alpha_1=10.5$). The other parameters are the same with those in Fig. \ref{fig:7}.
\label{fig:8}}
\end{figure}

As reported in previous works, chaos will not appear in weak driving and nonlinear coupling regime. This conclusion is still valid in our model. In Fig. \ref{fig:7}(a) and (b), we show that $\alpha_1$ always maintains a stable periodic oscillation and  evolves eventually to a limit cycle whose radius is small due to the weak driving. As contrast, the random-like evolution will appear in this system once the asymmetrical coupling exists, shown in Fig. \ref{fig:7}(c) and corresponding phase diagram is illustrated in (d) \cite{59}. The physical mechanism of such a chaos behavior in weak driving and coupling is similar with gain induced chaos in Ref. \cite{17}, that is, imaginary part of one eigenfrequency is positive (point A in Fig. \ref{fig:8}(a)), so that the variables of the system are amplified and the effective nonlinear coupling strength is enhanced indirectly. Therefore, as shown in Fig. \ref{fig:8}(b), the chaos will not appear once $\kappa_c$ passes the EP, which leads imaginary parts of eigenfrequency to be degenerate and become negative (point B in Fig. \ref{fig:8}(a)). We also explain that the spontaneous heating mechanism, in order to make $\kappa_{c\text{eff}}<\kappa_c$, is necessary for the induced chaos. Because the asymmetrical coupling is established through the cavity leakage in our model, $\kappa_{c\text{eff}}=\kappa_c$ will lead the corresponding intrinsic energy to
\begin{equation}
\begin{split}
E_{\pm}(\lambda_c)=-\Delta_{c\text{eff}}-i\dfrac{\kappa_c}{2}\pm\sqrt{\lambda_c(\lambda_c-\kappa_c)}.
\end{split}
\label{eq:chaosauida}
\end{equation}
Then Im$(E_+)_{max}=0$ with $\lambda_c=\kappa_c/2$ is obvious, as the dash line in Fig. \ref{fig:8}(a), meaning that imaginary parts of eigenfrequencies are always non-positive. In Fig. \ref{fig:8}(c), we show that the system returns back to the periodic evolution once $\kappa_{c\text{eff}}=\kappa_c$ although $\lambda_c$ is set the same with that in Fig. \ref{fig:7}(c) (point C in Fig. \ref{fig:8}(a)). Finally, by plotting the evolutions of Re$(\alpha_1)$ and $\vert\alpha_1\vert$ in Fig. \ref{fig:8}(d), we illustrate that the chaos is the sensitive dependence on the initial state, which is a key feature of chaotic evolutions \cite{60,61} 

\section{discussion and conclusion}
\label{discussion and conclusion}
In summary, we have found that a non-Hermitian interaction can also induce system to present a real spectrum. As an example, we have considered two quantum resonances with a non-Hermitian interaction and the  spectrum analyses show that there exists an EP, and the degenerate real parts (splitting imaginary part)  of the spectrum will become  splitting (degenerate) once some key parameters pass EP. This phenomenon is similar with that in gain-loss system, that is, an EP separates parameter interval into $\mathcal{PT}$BP and $\mathcal{PT}$SP. The non-Hermitian interaction considered is feasible with existing mature quantum optical devices. As a platform, we have also designed an optomechanical system to realize and observe dynamic properties of the non-Hermitian interaction. In particular, phase transition near the EP and gain induced chaos have been discussed.  The results indicate that our non-Hermitian system can be an excellent replacement for  $\mathcal{PT}$-symmetric system and it can avoid introducing any gains into the system. The interpretation of non-Hermitian interaction provide a platform for realizing parity-time symmetry devices and studying properties of non-Hermitian quantum mechanics.

\begin{acknowledgments}
All authors thank Jiong Cheng, Wenzhao Zhang, Xingyuan Zhang and Yang Zhang for the useful discussion. This research was supported by the National Natural Science Foundation of China (Grant No. 11574041, 11704026 and 11175033).
\end{acknowledgments}

\appendix
\section{Derivation of eliminating oscillator modes}
With the accuracy Langevin equations of system mean values in the main text, a set of zeroth order solutions of optical modes can be obtained as 
\begin{equation}
\begin{split}
{\alpha}_j(t)\simeq \alpha_j(0)\exp[(i\Delta _j-\dfrac{\kappa}{2})t],
\end{split}
\label{eq:A1}
\end{equation}
by neglecting all the interaction terms in weak coupling regime $\lambda$, $G$, $\kappa\ll\Delta_j$. Moreover, the form solutions of the oscillator modes are:
\begin{widetext}
\begin{equation}
\begin{split}
{\beta}_j(t)=&\beta_j(0)\exp[(-i\omega_m-\dfrac{\gamma_m}{2})t]+\int^t_0d\tau[-iG^{*}\alpha^*_j(\tau)]\exp[(-i\omega_m-\dfrac{\gamma_m}{2})(t-\tau)].
\end{split}
\label{eq:A2}
\end{equation}
Substituting Eq. \eqref{eq:A1} into Eq. \eqref{eq:A2} and finishing the integration in it, we have:
\begin{equation}
\begin{split}
{\beta}_j(t)=&\beta_j(0)\exp[(-i\omega_m-\dfrac{\gamma_m}{2})t]+\dfrac{2iG^*\alpha^*_j(0)\exp[(-i\Delta _j-\dfrac{\kappa}{2})t]}{2i(\Delta_j-\omega_m)-(\gamma_m-\kappa)}.
\end{split}
\label{eq:A3}
\end{equation}
In this expression, the integral lower bound is ignored. Under the conditions $\omega_m-\Delta_j\gg \vert G\vert$, $\alpha_j(0)\gg\beta_j(0)$ and $\gamma_m\gg\kappa$, the first term in above expression is a small quantity with high frequency and it will decay rapidly. By neglecting the first term, Eq. \eqref{eq:A3} becomes ${\beta}^*_j(t)\simeq\frac{2iG\alpha_j(t)}{2i(\Delta_j-\omega_m)+\gamma_m}$, and we have:
\begin{equation}
\begin{split}
&\dot{\alpha}_1=\left(i\Delta_1-\dfrac{\kappa}{2}\right)\alpha_1+\dfrac{2\vert G\vert^2\alpha_1}{2i(\Delta_1-\omega_m)+\gamma_m}+\lambda \alpha_2\\
&\dot{\alpha}_2=\left(i\Delta_2-\dfrac{\kappa}{2}\right)\alpha_2+\dfrac{2\vert G\vert^2\alpha_2}{2i(\Delta_2-\omega_m)+\gamma_m}-\lambda \alpha_1+\kappa\alpha_1,
\end{split}
\label{eq:A5}
\end{equation}
and it can be further simplified as
\begin{equation}
\begin{split}
&\dot{\alpha}_1=\left[i\left(\Delta_1-\dfrac{4\vert G\vert^2(\Delta_1-\omega_m)}{4(\Delta_1-\omega_m)^2+\gamma_m^2}\right)-\dfrac{1}{2}\left(\kappa-\dfrac{4\gamma_m\vert G\vert^2}{4(\Delta_1-\omega_m)^2+\gamma_m^2}\right)\right]\alpha_1+\lambda \alpha_2\\
&\dot{\alpha}_2=\left[i\left(\Delta_2-\dfrac{4\vert G\vert^2(\Delta_2-\omega_m)}{4(\Delta_2-\omega_m)^2+\gamma_m^2}\right)-\dfrac{1}{2}\left(\kappa-\dfrac{4\gamma_m\vert G\vert^2}{4(\Delta_2-\omega_m)^2+\gamma_m^2}\right)\right]\alpha_2-\lambda \alpha_1+\kappa\alpha_1.
\end{split}
\label{eq:A5}
\end{equation}
Eq. \eqref{eq:A5} is exactly the same with the Eq. \eqref{eq:effectmeavalueqle} in the main text.

\begin{thebibliography}{35}
\expandafter\ifx\csname
natexlab\endcsname\relax\def\natexlab#1{#1}\fi
\expandafter\ifx\csname bibnamefont\endcsname\relax
  \def\bibnamefont#1{#1}\fi
\expandafter\ifx\csname bibfnamefont\endcsname\relax
  \def\bibfnamefont#1{#1}\fi
\expandafter\ifx\csname citenamefont\endcsname\relax
  \def\citenamefont#1{#1}\fi
\expandafter\ifx\csname url\endcsname\relax
  \def\url#1{\texttt{#1}}\fi
\expandafter\ifx\csname
urlprefix\endcsname\relax\def\urlprefix{URL }\fi
\providecommand{\bibinfo}[2]{#2}
\providecommand{\eprint}[2][]{\url{#2}}

\bibitem{1} C. M. Bender and S. Boettcher Phys. Rev. Lett. \textbf{80}, 5243 (1998).

\bibitem{2} C. M. Bender, Rep. Prog. Phys. \textbf{70}, 947 (2007).

\bibitem{3} L. Xiao, X. Zhan, Z. H. Bian, K. K. Wang, X. Zhang, X. P. Wang, J. Li, K. Mochizuki, D. Kim, N. Kawakami, W. Yi, H. Obuse, B. C. Sanders, and P. Xue, Nat. Phys. (2017) doi:10.1038/nphys4204.

\bibitem{4} B. Peng, \c{S}. K. \"Ozdemir, F. Lei1, F. Monifi, M. Gianfreda, G. L. Long, S. Fan, F. Nori, C. M. Bender, and L. Yang, Nat. Phys. \textbf{10}, 394 (2014).

\bibitem{5} Y. Ashida, S. Furukawa, and M. Ueda, Nat. Commun. \textbf{8}, 15791 (2017).

\bibitem{6} H. Hodaei, M. A. Miri, M. Heinrich, D. N. Christodoulides, and M. Khajavikhan, Science  \textbf{346}, 975 (2014).

\bibitem{7} Z. Zhang, Y. Zhang, J. Sheng, L. Yang, M. A. Miri, D. N. Christodoulides, B. He, Y. Zhang, and M. Xiao
Phys. Rev. Lett. \textbf{117}, 123601 (2016).

\bibitem{8} Y. Aur\'egan and V. Pagneux, Phys. Rev. Lett. \textbf{118}, 174301 (2017).

\bibitem{9} A. Mostafazadeh, J. Math. Phys. \textbf{43}, 205 (2002).

\bibitem{10} A. Mostafazadeh, J. Math. Phys. \textbf{43}, 2814 (2002).

\bibitem{11} S. Dey, A. Fring, and L. Gouba, J. Phys. A: Math. Theor. \textbf{48}, 40FT01 (2015).

\bibitem{12} E. M. Graefe, S. Mudute-Ndumbe, and M. Taylor, J. Phys. A: Math. Theor. \textbf{48}, 38FT02 (2015).

\bibitem{add8} R. B. B. Santos and V. R. da Silva, Mod. Phys. Lett. B, \textbf{28}, 1450223 (2014).

\bibitem{13} Z. Lin, H. Ramezani, T. Eichelkraut, T. Kottos, H. Cao, and D. N. Christodoulides, Phys. Rev. Lett.  \textbf{106}, 213901 (2011).

\bibitem{14} X. Luo, J. Huang, H. Zhong, X. Qin, Q. Xie, Y. S. Kivshar, and C. Lee, Phys. Rev. Lett. \textbf{110}, 243902 (2013).

\bibitem{15} C. Hang, G. Huang, and V. V. Konotop, Phys. Rev. Lett. \textbf{110}, 083604 (2013).

\bibitem{16} J. H. Wu, M. Artoni, and G. C. La Rocca, Phys. Rev. A \textbf{91}, 033811(2015).

\bibitem{17} X. Y. L\"u, H. Jing, J. Y. Ma, and Y. Wu, Phys. Rev. Lett. \textbf{114}, 253601 (2015).

\bibitem{18} H. Jing, S. K. \"Ozdemir, X. Y. L\"u, J. Zhang, L. Yang, and F. Nori, Phys. Rev. Lett. \textbf{113}, 053604 (2014).

\bibitem{19} T. E. Lee, Phys. Rev. Lett. \textbf{116}, 133903 (2016).

\bibitem{20} D. Leykam, K. Y. Bliokh, C. Huang, Y. D. Chong, and F. Nori, Phys. Rev. Lett. \textbf{118} 040401 (2017).

\bibitem{21} K. V Kepesidis, T. J Milburn, J. Huber, K. G Makris, S. Rotter, and P. Rabl, New J. Phys. \textbf{18}  095003 (2016).

\bibitem{22} Z. P. Liu, J. Zhang, \c{S}. K. \"Ozdemir, B. Peng, H. Jing, X. Y. L\"u, C. W. Li, L. Yang, F. Nori, and Y. X. Liu, Phys. Rev. Lett. \textbf{117}, 110802 (2016).

\bibitem{23} J. Li, R. Yu, and Y. Wu, Phys. Rev. A \textbf{92}, 053837 (2015).

\bibitem{24} J. Li, X. Zhan, C. Ding, D. Zhang, and Y. Wu, Phys. Rev. A \textbf{92}, 043830  (2015).

\bibitem{25}  X. Y. Zhang, Y. Q. Guo, P. Pei, and X. X. Yi, Phys. Rev. A \textbf{95} 063825 (2017).

\bibitem{26} M. Klett, H. Cartarius, D. Dast, J. Main, and G. Wunner, Phys. Rev. A \textbf{95}, 053626 (2017).

\bibitem{27} T. J. Milburn, J. Doppler, C. A. Holmes, S. Portolan, S. Rotter, and Peter Rabl, Phys. Rev. A \textbf{92}, 052124 (2015).

\bibitem{28} H. Menke and M. M. Hirschmann, Phys. Rev. B \textbf{95}, 174506 (2017).

\bibitem{29} J. S. Tang, Y. T. Wang, S. Yu, D. Y. He, J. S. Xu, B. H. Liu, G. Chen, Y. N. Sun, K. Sun, Y. J. Han, C. F. Li, and G. C. Guo, Nat. Photon. \textbf{10}, 642 (2016).

\bibitem{30} G. S. Agarwal and Kenan Qu, Phys. Rev. A \textbf{85}, 031802(R)  (2012).

\bibitem{31} W. L. Li, C. Li, and H. S. Song, Phys. Rev. A \textbf{95}, 023827 (2017).

\bibitem{32} J. Li, J. Li, Q. Xiao, and Y. Wu, Phys. Rev. A \textbf{93}, 063814 (2016).

\bibitem{33} X, Zhou and Y. D. Chong, Opt. Express. \textbf{24} 6916 (2016).

\bibitem{34} Y. Jiao, H. L\"u, J. Qian, and Y. Li and H. Jing, New J. Phys. \textbf{18},  083034 (2016).

\bibitem{35} W. L. Li, Y. F. Jiang, C. Li, and H. S. Song, Sci. Rep. \textbf{6}, 31095 (2016).

\bibitem{add9} Y. C. Lee, J. Liu, Y. L. Chuang, M. H. Hsieh, and R. K. Lee, Phys. Rev. A \textbf{92}, 053815 (2015).

\bibitem{add10} R. B. B. Santos, EPL \textbf{100}, 24005 (2012).

\bibitem{36} C. M. Bender, M. Gianfreda, \c{S}. K. \"Ozdemir, B. Peng, and L. Yang, Phys. Rev. A \textbf{88}, 062111 (2013).

\bibitem{37} X. W. Xu, Y. X. Liu, C. P. Sun, and Y. Li, Phys. Rev. A \textbf{92}, 013852 (2015).

\bibitem{add3} B. He, L. Yang, Z. Zhang, and M. Xiao, Phys. Rev. A \textbf{91}, 033830 (2015).

\bibitem{38} B. P. Mandal, B. K. Mourya, and R. K. Yadav, Phys. Lett. A \textbf{377} 1043 (2013).

\bibitem{39} C. M. Bender, B. K. Berntson, D. Parker, E. Samuel, Am. J. Phys. \textbf{81}, 173 (2013).

\bibitem{40} C. T. West, T. Kottos, and T. Prosen, Phys. Rev. Lett. \textbf{104}, 054102 (2010).

\bibitem{41} C. W. Gardiner, Phys. Rev. Lett. \textbf{70}, 2269 (1993).

\bibitem{42} J. I. Cirac, P. Zoller, H. J. Kimble, and H. Mabuchi, Phys. Rev. Lett. \textbf{78}, 3221 (1997).

\bibitem{43} S. J. vanEnk, J. I. Cirac, and P. Zoller, Phys. Rev. Lett. \textbf{78}, 4293 (1997).

\bibitem{44} S. Clark, A. Peng, M. Gu, and S. Parkins, Phys. Rev. Lett. \textbf{91}, 177901 (2003). 

\bibitem{45} B. Vermersch, P. O. Guimond, H. Pichler, and P. Zoller, Phys. Rev. Lett. \textbf{118}, 133601 (2017).

\bibitem{46} Z. L. Xiang, M. Zhang, L. Jiang, and P. Rabl, Phys. Rev. X \textbf{7}, 011035 (2017).

\bibitem{47} K. Stannigel, P. Rabl, A. S. S\o{}rensen, P. Zoller, and M. D. Lukin, Phys. Rev. Lett. \textbf{105}, 220501 (2010).

\bibitem{48} E. A. Sete and H. Eleuch, Phys. Rev. A \textbf{91}, 032309 (2015).

\bibitem{49} M. Aspelmeyer, T. J. Kippenberg, and F. Marquardt, Rev. Mod. Phys. \textbf{86}, 1391 (2014).

\bibitem{50} F. Marquardt and S. M. Girvin, Physics \textbf{2}, 40 (2009).

\bibitem{51} Y. C. Liu, Y.  F. Xiao, X. Luan, and C. W. Wong, Phys. Rev. Lett. \textbf{110}, 153606 (2013). 

\bibitem{add1} W. Z. Zhang, J. Cheng, W. D. Li, and L. Zhou, Phys. Rev. A \textbf{93}, 063853 (2016). 

\bibitem{52} C. W. Gardiner and P. Zoller, \textit{Quantum Noise} (Springer, Berlin, 2000).

\bibitem{add6} Here, $\langle...\rangle$ implies taking the expectation value with respect to the density matrix of the quantum system.

\bibitem{add4} Y. C. Liu, X. Luan, H. K. Li, Q. Gong, C. W. Wong, and Y. F. Xiao, Phys. Rev. Lett. \textbf{112}, 213602 (2014). 

\bibitem{add5} T. Y. Chen, W. Z. Zhang, R. Z. Fang, C. Z. Hang, and L. Zhou, Opt. Express \textbf{25}, 10779 (2017).

\bibitem{add2} Here $\alpha_j$ and $\alpha'_j$ denote results basing on the accuracy Langevin equations \eqref{eq:meavalueqle} and effective Langevin equations \eqref{eq:effectmeavalueqle},  respectively.

\bibitem{53} A. Mari and J. Eisert, Phys. Rev. Lett. \textbf{103}, 213603 (2009).

\bibitem{59} L. L\"u, C. R. Li, G. Li, A. Sun, Z. Yan, T. T. Rong, Y. Gao, Commun. Nonlinear Sci. Numer. Simulat. \textbf{47}, 267 (2017).

\bibitem{54} W. L. Li, W. Z. Zhang, C. Li, and H .S. Song, Phys. Rev. E \textbf{96}, 012211 (2017). 

\bibitem{55} Here $\delta o=o-\bar{o}$ and $\bar{o}=\Delta t^{-1}\int^{t+\Delta t}_{\Delta t} o(t)dt$.

\bibitem{56} J .Wong, J. Math. Phys. \textbf{8}, 2039 (1967). 

\bibitem{57} C. P. Sun, Phys. Scr. \textbf{48}, 393 (1993).

\bibitem{58} G. Manzano, F. Galve, G. L. Giorgi, E. Hern\'andez-Garc\'ia, and R. Zambrini, Sci. Rep. \textbf{3}, 1439 (2013).

\bibitem{60} L. L\"u, C. R. Li, L. S. Chen, G. N. Zhao, Nonlinear Dyn. \textbf{86}, 655 (2016).
 
\bibitem{61} L. L\"u, C. R. Li, G. N. Zhao,  Sci. Sin.-Phys. Mech. Astron. \textbf{47}, 080501 (2017).

\end{thebibliography}
\end{widetext}

\end{document}